\documentclass[conference]{IEEEtran}

\usepackage{subcaption}     
\usepackage{makecell}
\usepackage{graphicx}
\usepackage{latexsym}
\usepackage{amsxtra} 
\usepackage{amsmath,amssymb,amsfonts,stmaryrd}
\usepackage{marvosym}
\usepackage{scalefnt}
\usepackage{color,tabularx}
\usepackage{makecell}
\usepackage{booktabs,multicol,multirow}
\usepackage{algorithm}
\usepackage{comment}
\usepackage{mathtools}
\usepackage[font=small]{caption}
\usepackage{url}
\usepackage{soul}
\usepackage{cite}
\usepackage[dvipsnames]{xcolor}
\usepackage[noend]{algpseudocode}
\usepackage[hidelinks]{hyperref}
\usepackage{dblfloatfix}  
\usepackage[capitalise]{cleveref}
\usepackage{pifont}
\usepackage{soul}
\usepackage{threeparttable}

\usepackage[  round-mode = places, round-precision = 2]{siunitx}

\usepackage{adjustbox}
\usepackage{tabularx}
\usepackage{array}
\newcolumntype{P}[1]{>{\centering\arraybackslash}m{#1}}

\usepackage[a4paper, total={184mm,239mm}]{geometry}
\def\BibTeX{{\rm B\kern-.05em{\sc i\kern-.025em b}\kern-.08em
    T\kern-.1667em\lower.7ex\hbox{E}\kern-.125emX}}

\makeatletter
\newcommand*\titleheader[1]{\gdef\@titleheader{#1}}
\AtBeginDocument{%
  \let\st@red@title\@title
  \def\@title{%
    \bgroup\normalfont\normalsize\centering\@titleheader\par\egroup
    \vskip1ex\st@red@title}
}
\makeatother

\title{Computing with Printed and Flexible Electronics}

\author{
    \IEEEauthorblockN{
        Mehdi B. Tahoori\IEEEauthorrefmark{1},
        Emre Ozer\IEEEauthorrefmark{2},
        Georgios Zervakis\IEEEauthorrefmark{4},
        Konstantinos Balaskas\IEEEauthorrefmark{4}, and
        Priyanjana Pal\IEEEauthorrefmark{1}
    }
    \IEEEauthorblockA{
        \IEEEauthorrefmark{1}Karlsruhe Institute of Technology, Karlsruhe, Germany,\\
        \IEEEauthorrefmark{2}Pragmatic Semiconductor Ltd, Cambridge, UK,\\
        \IEEEauthorrefmark{4}University of Patras, Patras, Greece
    }
    \IEEEauthorblockA{
        \IEEEauthorrefmark{1}\{mehdi.tahoori, priyanjana.pal\}@kit.edu,
        \IEEEauthorrefmark{2}eozer@pragmaticsemi.com,
        \IEEEauthorrefmark{4}\{zervakis,kompalas\}@ceid.upatras.gr
    }
}
\titleheader{\vspace{-25pt}To appear at the 30th IEEE European Test Symposium May 26-30, 2025, Tallinn, Estonia. \textsuperscript{\textcopyright}IEEE}

\begin{document}
\bstctlcite{IEEEexample:BSTcontrol} 
\maketitle
\begin{abstract}
Printed and flexible electronics (PFE) have emerged as the ubiquitous solution for application domains at the extreme edge, where the demands for low manufacturing and operational cost cannot be met by silicon-based computing.
Built on mechanically flexible substrates, printed and flexible devices offer unparalleled advantages in terms of form factor, bio-compatibility and sustainability, making them ideal for emerging and uncharted applications, such as wearable healthcare products or fast-moving consumer goods.
Their desirable attributes stem from specialized fabrication technologies, e.g., Pragmatic's FlexIC, where advancements like ultra-thin substrates and specialized printing methods expand their hardware efficiency, and enable penetration to previously unexplored application domains.
In recent years, significant focus has been on machine learning (ML) circuits for resource-constrained on-sensor and near-sensor processing, both in the digital and analog domains, as they meet the requirements of target applications by PFE.
Despite their advancements, challenges like reliability, device integration and efficient memory design are still prevalent in PFE, spawning several research efforts towards cross-layer optimization and co-design, whilst showing promise for advancing printed and flexible electronics to new domains.
\end{abstract}

\section{Introduction}\label{sec:intro}

The continuous advancement of computing systems has made them an integral part of daily life, expanding in scale and impact with each technological breakthrough.
For over five decades, silicon (Si)-based microprocessors have been the foundation of semiconductor technology, consistently improving in performance, area efficiency, power consumption, and cost.
However, despite their relatively low unit cost of a few dollars, Si microprocessors face inherent evolutionary limitations that make them unsuitable for many emerging applications.
These include fast-moving consumer goods (e.g., smart labels and packaging), wearable healthcare devices (e.g., smart patches and dressings), disposable medical implantables (e.g., neural interfaces), and diagnostic test strips (e.g., lateral flow tests).
Such applications require not only extremely low costs but also flexibility, stretchability, and compact form factors--characteristics that conventional Si technology struggles to provide.

Printed and flexible electronics (PFE) emerge as a solution to address these limitations, offering unparalleled advantages in cost and form factor, to enable these emerging application domains.
Silicon manufacturing comes with major operational, testing, and compliance expenses, whereas flexible integrated circuits (FlexICs) can be fabricated in smaller distributed facilities, with cheaper and even portable equipment, and without the need for protective packaging, thanks to their physical attributes~\cite{Bleier:ISCA:2020:printedmicro}.
Moreover, a key feature of PFE is sustainability.
For example, the FlexIC process of Pragmatic Semiconductor, which is based on Indium Gallium Zinc Oxide (IGZO) thin-film transistors (TFTs), enables rapid production cycles while dramatically reducing environmental impacts--achieving a significant reduction in water usage, energy consumption, and carbon footprint~\cite{ozer:nature2024:bendableRiscV}.
Furthermore, their conformal, porous, and stretchable properties make printed and flexible technologies ideal for applications requiring physical flexibility and lightweight designs.

While printed and flexible circuits enable the development of ultra-low-cost and conformal hardware, they face significant challenges in achieving large-scale integration.  
These challenges originate from their relatively large feature sizes, limited device count and integration density, and high device latencies—several orders of magnitude below that of silicon-based VLSI~\cite{Henkel:ICCAD2022:expedition}, mainly stemming from their low-cost fabrication processes. 
Printed electronics technology is significantly slower (in the order of Hz) but enables in-situ tuning and point-of-use customization, while flexible electronics provide better performance (kHz range) and significantly higher integration density.
Despite these limitations, PFE can meet the requirements of certain applications where performance and precision demands are minimal, such as circuits operating at low sampling rates (a few Hz) and limited bit precision~\cite{Henkel:ICCAD2022:expedition}.
As such, they are emerging as key enablers for computing in domains where conventional silicon systems have not yet achieved significant penetration. 

Additionally, a key technological limitation is the lack of a robust p-type device in metal oxide, or a robust n-type one in organic electronic technologies.
Consequently, printed and flexible circuits often rely on resistive-load configurations, such as resistor n-MOS designs.  
This poses significant constraints in low-power designs, as the use of resistive loads leads to higher power consumption, increased area requirements, and greater variability between devices compared to CMOS technology. 

PFE may also be susceptible to aging-related degradation due to thermal stress over time, as well as process variations caused by manufacturing inaccuracies such as non-uniform device geometries and ink dispersion in printed electronics~\cite{Zhao:IDP:2023:variability_pe, Rasheed:IEEE-TED:2018:variability_model}.  
Moreover, while they can resist mechanical stress, such as bending, without needing additional chip packaging like Si dies, they need to be validated, tested, and ensure operation under varying mechanical stress conditions like tensile and compressive modes~\cite{ozer:nature2024:bendableRiscV}.
These limitations also present substantial challenges in the design, reliability, and testing of printed and flexible circuits, thus necessitating novel {\em Design for Reliability (DfR)} solutions integrated in their design methodologies and complemented with post-manufacturing calibration and optimization. 

Extensive research efforts have been dedicated to address the aforementioned challenges.
This work presents the state-of-the-art in PFE, offering an overview from technology fundamentals and advancements, to practical applications, to novel design methodologies.
\cref{sec:tech} provides detailed information on emerging technologies and recent technological milestones.
These include  low-voltage, high-mobility Electrolyte-Gated FETs (EGFETs) enabling battery-powered operation~\cite{Bleier:ISCA:2020:printedmicro}, IGZO-TFTs on polyimide substrates supporting large-scale non-silicon microprocessors~\cite{ozer:nature2024:bendableRiscV}, and Pragmatic’s FlexIC Gen3 platform showcasing mechanically robust submicron IGZO-TFTs on ultra-thin polyimide~\cite{flexic_gen3}.
\cref{sec:app} focuses on application domains in which PFE insert intelligence at sub-cent costs, enabling affordable yet efficient computation at the extreme edge.
In \cref{sec:arch}, several state-of-the-art design strategies for printed/flexible circuits are detailed, spanning from the digital to the analog domain. 
We focus on both general-purpose computing elements (e.g., flexible RISC-V microprocessors~\cite{ozer:nature2024:bendableRiscV}), as well as domain-specific circuits which adopt high degrees of customization through bespoke and/or approximate designs.
Bespoke circuits specifically leverage the low non-recurring engineering (NRE) costs of such technologies and tailor the circuit implementation to the application at hand, achieving impressive reductions in area, power and energy consumption, effectively enabling battery-powered operation.
\cref{sec:reliability} presents an analysis of current efforts towards improving the reliability and robustness of printed and flexible devices and circuits, through fault sensitivity analysis, dedicated test generation methods, and robustness-aware training.

\begin{table*}[t]

    \centering
    \caption{Comparison of natively flexible general-purpose processors fabricated and validated}
    \label{tab:flexmicroprocessor}
    \setlength\tabcolsep{2pt}
    \renewcommand{\arraystretch}{1.2}
    
     \begin{tabularx}{\textwidth}{|X|P{1.4cm}|P{1.4cm}|P{1.3cm}|P{1.55cm}|P{1.35cm}|P{1.35cm}|P{1.45cm}|P{1.4cm}|P{1.2cm}|P{1.4cm}|P{1.25cm}|} \hline
      \textbf{Features} & \textbf{PlasticCPU} \cite{takayama04} & \textbf{RFCPU}\newline\cite{dembo05} & \textbf{Asynch. \textmu proc.} \cite{karaki05} & \textbf{UHF}\newline\textbf{RFCPU} \cite{kurokawa08} &  \textbf{Organic \textmu proc.} \cite{myny12} & \textbf{Hybrid \textmu proc.} \cite{myny14}  & \textbf {PlasticARM} \cite{biggs21} & \textbf{Flex6502-1} \cite{celiker22} & \textbf{FlexiCore} \cite {bleier22} & \textbf{Flex6502-2} \cite{celiker24} & \textbf{Flex-RV} \cite{ozer:nature2024:bendableRiscV}\\ \hline
      \textbf{Data\newline Bit-width}  & 8  &  8 &  8 &  8 &  8 & 8 & 32 &  8 & 4 and 8 & 8 & 32\\ \hline
      \textbf{ISA}  & Custom & Custom & Custom & Custom & Custom & Custom & ARM & 6502 & Custom & 6502 & RISC-V\\ \hline
      \textbf{S: Synch. or A:Asynch.}  & S & S & A & S & S & S & S & S &  S & S & S\\ \hline
      \textbf{CPU or SoC}  & CPU &  SoC (ROM, RAM \& RF Interface) & CPU & SoC (ROM, RAM \& RF Interface) &  CPU & SoC (CPU \& ROM) & SoC (RAM, ROM \& Peripherals) & CPU & CPU & CPU & SoC (Peripherals \& ML Accelerator)\\ \hline
      \textbf{TFT Type}  & Poly-silicon &  Poly-silicon & LTPS & Poly-silicon & Organic & Organic + Metal-oxide & Metal-oxide & Metal-oxide & Metal-oxide & LTPS & Metal-oxide\\ \hline
      \textbf{Technology Node (\textmu)}  & 1.2 &  1 & 4 & 0.8 & 5 & 5 & 0.8 & 0.8 & 0.8 & 3 & 0.6\\ \hline
      \textbf{Logic Type}  & CMOS & CMOS & CMOS & CMOS & CMOS & Pseudo-CMOS & Resistive Load & Pseudo-CMOS & Resistive Load & CMOS & Resistive Load\\ \hline
      \textbf{Supply Voltage (V)}  & 3.3 & 1.8 & 3.5-7 & 1.8 &  10 & 12 & 3-4.5 & (2 \& 3) or (3 \& 6) & 4.5 & 9 & 3 \\ \hline
      \textbf{Clock Frequency (kHz)}  & 10,000 &  3,390 & 30@3.5V and 500@5V & 1120 & 0.04 & 2.1 & 29-40 &  10-71 & 12.5 & 454.5 & 60 \\ \hline
      \textbf{Core Area (mm\textsuperscript{2})}  & Not reported &  196  & 156.25 & 93.45 & 33,700 & 22,600 & 59.2 & 24.91 & 5.56 \& 6.06 & 75.33 & 17.5\\ \hline
      \textbf{Number of Devices}  & Not reported &  71,000  & 32,000  & 133,000  &  3381 & 3504 & 56,340 &  16,392 & 2,104 \& 2,335 & 12,628 & $\sim$38,000\\ \hline
      \textbf{NAND2-equivalent Gatecount}  & Not reported &  17,750\newline(Estimated) & 8,000\newline(Estimated) & 33,000\newline(Estimated) &  1127 & 876  & 18,334 & 2732 & 801 & 3200\newline(Estimated) & 12,596 \\ \hline
      \textbf{Power (mW)}  &  Not reported & 4.14 & 0.725@5V & 0.81 & 0.092 & Not reported &  21 & 11.6-134.9 & 1.8 \& 2.4 & 15.3 & 5.8\\ \hline
      \textbf{Year}  & 2004 &  2005 & 2005 & 2008 &  2012 & 2014 & 2021 & 2022 & 2022 & 2024 & 2024\\ \hline
    \end{tabularx}
    
\end{table*}
\section{Technology and Fabrication}\label{sec:tech}
\begin{figure}[t!]
    \centering
    \includegraphics[width =\columnwidth]{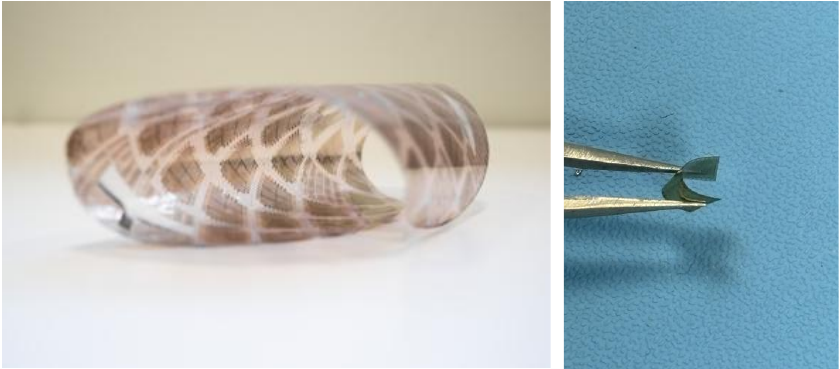}
    \caption{A flexible wafer on a polyimide substrate, and a FlexIC held in a tweezer}
    \label{fig:flexic}
\end{figure}

\subsection{Pragmatic FlexIC Technology}\label{subsec:flexic}
Pragmatic's FlexIC technology uses an advanced process to fabricate physically flexible integrated circuits on a 200/300 mm polyimide wafers (see Fig. \ref{fig:flexic}). 
The Gen3 FlexIC \cite{flexic_gen3} is the latest platform that utilizes resistive n-type metal-oxide TFT technology, based on IGZO for production of FlexICs, by running several sequences of material deposition, patterning and etching.
The Gen3 platform supports TFTs with a minimum channel length of 0.6 µm, and integrates key components such as resistors and capacitors.
The polyimide substrate is ultra thin, with a thickness of less than 30 µm.
FlexICs have demonstrated mechanical robustness with a radius of curvature as small as 3 mm without any damage to circuitry.

\begin{figure}[t!]
    \centering
    \includegraphics[width=\columnwidth]{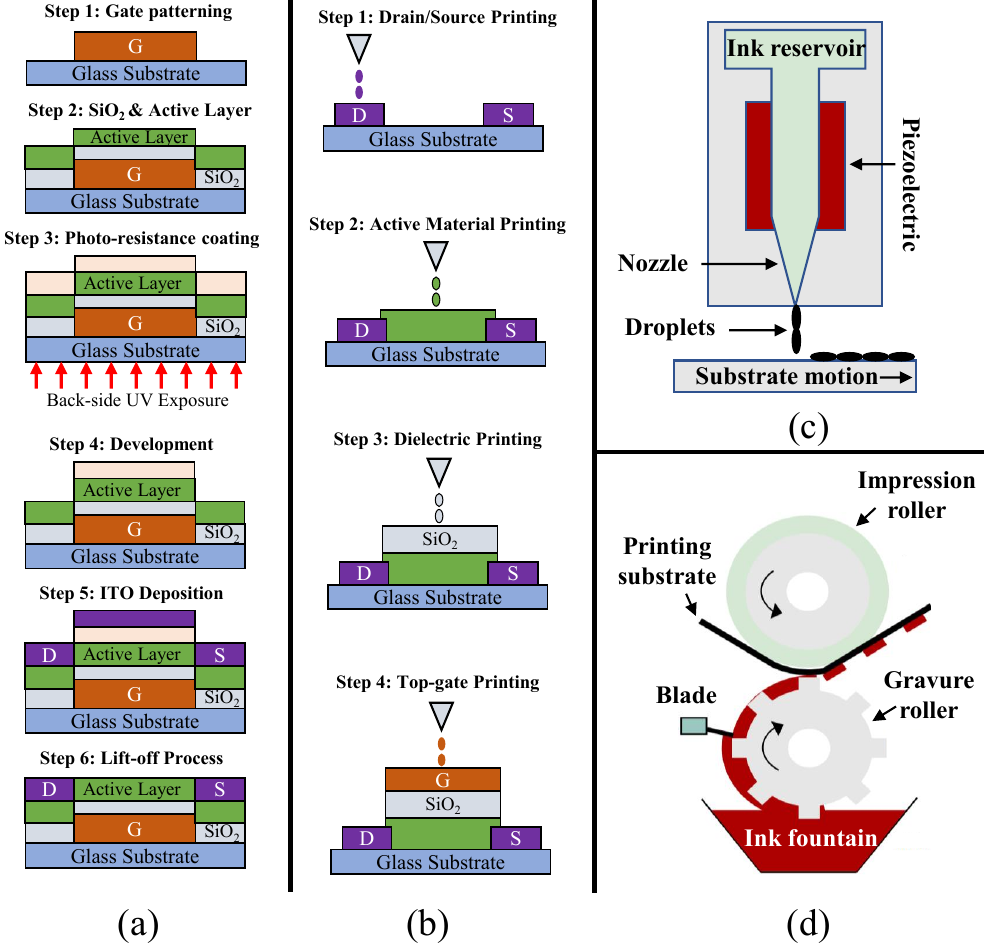}
    \caption{Illustration of printed manufacturing techniques. (a) Subtractive and (b) additive processes, (c) inkjet and (d) gravure printing.}
    \label{fig:pe_tech}
\end{figure}

\subsection{Printed Electronics}\label{subsec:printed}
Printed electronics (PE) technology denotes a set of fabrication methods that leverage various printing techniques, including inkjet, screen, and gravure printing~\cite{cui2016printed}. 
Optimal device performance in PE is traditionally achieved using vacuum-deposited highly purified molecular substrates. However, recent advancements in solution-based fabrication methods, such as spin-coating and inkjet printing, have attracted significant attention due to their potential for enhancing manufacturing efficiency and significantly reducing production costs. Printing technologies generally fall into two primary categories, illustrated in \cref{fig:pe_tech}. First, replication printing techniques, such as gravure printing (\cref{fig:pe_tech}(d)), are particularly suited for high-throughput, large-scale production. Second, jet printing methods, like aerosol or inkjet printing (\cref{fig:pe_tech}(c)), cater to customized fabrication of electronic circuits in smaller, tailored batches.
Furthermore, manufacturing approaches in PE can be classified as either subtractive or additive, shown in \cref{fig:pe_tech} (a) and (b), respectively. Subtractive manufacturing involves cycles of material deposition followed by selective etching, typically demanding specialized equipment and thus incurring higher costs. On the other hand, additive manufacturing sequentially deposits materials layer by layer to construct electronic components, a method clearly illustrated by inkjet printing. Although additive methods produce components with lower resolution and higher variability, their cost-effectiveness makes them highly appealing for various applications in PE.
The minimal equipment requirements and the simplicity of mask-less additive processes allow for the fabrication of extremely low-cost electronic circuits--sometimes costing less than a cent--at reduced development time~\cite{chang2017circuits}.

In the respective literature, the printed EGFET technology~\cite{Bleier:ISCA:2020:printedmicro} has attracted major interest.
EGFET devices boast favorable mobility characteristics and operate at low supply voltages, as they are capable of functioning below $1$V and even as low as $0.6$V~\cite{Marques:Materials:2019}, aligning well with battery-powered printed applications.


\section{Applications}\label{sec:app}

PFE are specifically suitable for applications at the extreme edge for on-sensor and near-sensor processing, which is defined as the domain where the electronics deployed in an end device cannot be based on conventional Si technology because of one or more of these constraints: form factor, cost, conformability, biocompatibility and user comfort.  An extreme-edge end device will need to process sensor data to extract knowledge, which typically involves predictions such as pattern recognition or classification.

Good examples of the extreme edge applications are logistics, fast-moving consumer goods (FMCG), healthcare wearables, implantable/ingestibles, textiles, agricultural and environmental monitoring. Smart packaging in the FMCG domain involves very cost-sensitive applications, where embedded electronics cannot cost more than a few cents. Making predictions is important for many FMCG applications, such as a smart package embedded with a flexible chip which can predict food freshness. Another example is an ECG patch or a smart dressing in healthcare domain, which requires conformable and comfortable electronics for patients. Such a patch or dressing--embedded with a flexible chip--can predict events like arrythmia/AF in the ECG patch, or healed wounds in the smart dressing.
In the agricultural monitoring domain, a patch embedded with a flexible chip can be wrapped around a plant to monitor plant growth and predict growth anomalies.

\section{Computing Architecture}\label{sec:arch}
\subsection{General-purpose Processors}\label{subsec:micpro}

Natively flexible general-purpose processors can be grouped into categories, based on the instruction set architecture (ISA) they follow, the utilized computational bitwidth and the semiconductor material with which they are developed.
Early processors were 8-bit CPUs \cite{takayama04, dembo05, karaki05, kurokawa08} with proprietary ISA, and developed using low-temperature poly-silicon (LTPS) TFT technology that has a high manufacturing cost and poor lateral scalability. 
An 8-bit arithmetic logic unit (ALU) with a print-programmable ROM was developed with metal-oxide TFTs, and was fabricated on polyimide~\cite{myny12, myny14}. 
Flex6502~\cite{celiker22} was an 8-bit processor in 6502 ISA-implemented and fabricated using IGZO-based FlexIC technology. Also, a low-temperature poly-silicon (LTPS) version of Flex6502 was fabricated and demonstrated \cite{celiker24}. FlexiCores \cite {bleier22} performed yield analysis of 4-bit and 8-bit processors in FlexIC technology. 

PlasticARM \cite{biggs21} was the first 32-bit processor developed using flexible electronics technology. It was a system-on-a-chip (SoC) comprised of a 32-bit ARM CPU derived from the Arm® Cortex-M0+® processor supporting the Armv6-M architecture.
It also integrated CPU peripherals, a ROM and a small latch-based RAM, all fabricated as a FlexIC.
Flex-RV~\cite{ozer:nature2024:bendableRiscV} was the first 32-bit RISC-V microprocessor in FlexIC, an important milestone in building ultralow-cost bendable computing architectures. 
Fabricated using 0.6 µm IGZO TFTs on a 30 µm polyimide substrate, Flex-RV is ultrathin, bendable, and sub-dollar in cost, making it ideal for low-cost, flexible electronics.
It also supports ML workloads via an integrated programmable hardware accelerator, and runs at up to 60 kHz while consuming under 6 mW, maintaining functionality even under a 3 mm bending radius.

Table \ref{tab:flexmicroprocessor} shows the features and properties of the natively flexible general-purpose processors that have been fabricated and validated. The table does not include simulation-based printed and flexible processor studies that have not been fabricated and validated such as \cite{chang17, Bleier:ISCA:2020:printedmicro}.


\subsection{Domain-specific Digital Processing Elements}\label{subsec:asic}

Because of the highly embedded nature and short lifespan of many extreme edge applications, the development of domain-specific processing engines in PFE has emerged as the de-facto solution to address the inherent limitations in these technologies~\cite{Ozer2019Bespoke}.
Available resources in printed and flexible circuits are very limited, typically constrained to only a few thousand gates~\cite{Iordanou2024,Mubarik:MICRO:2020:printedml,Bleier:ISCA:2020:printedmicro}.
Likewise, printed batteries can supply only a few tens of milliwatts of power~\cite{Mubarik:MICRO:2020:printedml}, which is sufficient to drive, at most, a few thousand gates.

By leveraging the extremely low fabrication and NRE costs in PFE, high degrees of customization can be embedded within the circuit implementation, thus significantly reducing overheads associated with general-purpose functionality.
Such customization levels are largely impractical in silicon technologies, due to the high costs of silicon process technology-including masksets and packaging costs.

In PFE, specialization is mainly enabled through alternative computing paradigms such as bespoke and approximate computing~\cite{Armeniakos:DATE2022:axml, Armeniakos:TCAD2023:cross,Armeniakos:TC2023:codesign,Afentaki:DATE2024:gatrain,Afentaki:ICCAD23:hollistic,Weller:2021:printed_stoch,Kokkinis:DATE2023,Armeniakos:DATE2024:dt,Afentaki:ESL2024,Mrazek:ICCAD2024:tnn}.
Bespoke circuits are highly customized and prioritize hardware efficiency over generalization capabilities~\cite{Ozer2019Bespoke,Ozer:DFLEPS2020, Ozer:Nature:2020,Iordanou2024, Bleier:ISCA:2020:printedmicro,Mubarik:MICRO:2020:printedml}.
They also enable more compact designs with a reduced gate count, leading to lower marginal costs.
Therefore, the bespoke design paradigm has not only gained significant traction in the development of complex systems, such as ML classifiers, for printed and flexible technologies, but has also emerged as the design standard due to its exceptionally high hardware efficiency.

\begin{figure}[t!]
    \centering
    \includegraphics[width=0.7\linewidth]{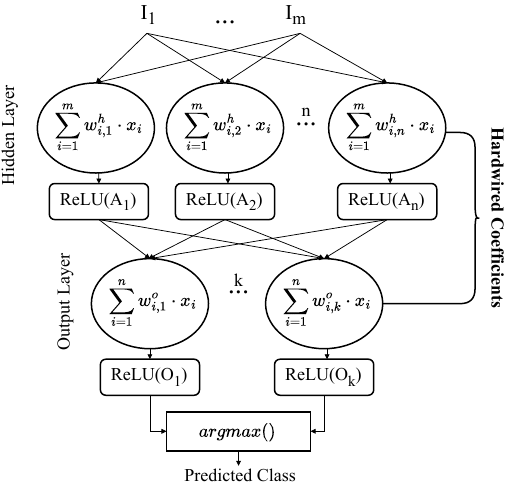}
    \caption{Block diagram of fully-parallel bespoke MLP.}
    \label{fig:bespoke_mlp}
\end{figure}

\subsubsection{Bespoke ML Classifiers}
Printed and flexible ML represent an extreme subset within the TinyML domain where advancements have mainly been driven by bespoke ML circuits. 
Bespoke classifiers are tailored to a specific model, trained for a given application and dataset.
For instance, a bespoke ML engine would hardwire the trained parameters directly into the circuit implementation. In fact, the first flexible ML chips are designed and fabricated as FlexICs, following the bespoke design principal~\cite{Ozer:Nature:2020, Ozer:DFLEPS2020}.
Specifically, \cite{Ozer:Nature:2020} implements a fully-parallel bespoke classification system as a FlexIC, whilst \cite{Ozer:DFLEPS2020} is a flexible and hardwired binary neural network (BNN). Recently, arrWNN \cite{velu24} has demonstrated a weightless neural network (WNN) implemented and fabricated as a FlexIC to detect arrhythmia events.
An overview of existing printed and flexible bespoke ML classifiers is presented in Table~\ref{tab:table_bespokeclf}.

Studies like~\cite{Armeniakos:DATE2022:axml,Armeniakos:TCAD2023:cross,Balaskas:ISQED2022:axDT} have demonstrated that bespoke arithmetic units--such as multipliers with a hardwired multiplicand set to a specific constant--can be, on average, $5$x smaller than their conventional counterparts.
Additionally, embedding constants directly into the RTL description allows logic synthesis tools to propagate constants and further optimize logic downstream~\cite{Mubarik:MICRO:2020:printedml}.

Most state-of-the-art domain-specific printed and flexible ML ASICs adopt fully parallel bespoke architectures, where model parameters are hardwired directly into the circuit design. These implementations are entirely combinational, meaning that the classifier’s outputs update instantly in response to input changes.
For instance, in a neural network~\cite{Kokkinis:TC2025,Armeniakos:TC2023:codesign,Mrazek:ICCAD2024:tnn}, each neuron is assigned dedicated hardware, with each weight instantiating a separate multiplier, and all neurons operate in parallel. \cref{fig:bespoke_mlp} presents a block diagram of a purely-combinational fully parallel bespoke multi-layer perceptron (MLP).

Although some studies have explored folded implementations~\cite{Sertaridis:ISCAS2025,Besias:DATE2025,Saglam:ASPDAC2025,Mubarik:MICRO:2020:printedml}, sequential engines remain scarce for these technologies.
While a flip-flop in CMOS technology occupies the area of approximately four NAND gates, it is equivalent to six NAND gates in EGFET technology--incurring a $50$\% overhead compared to CMOS.
Moreover, memory options in PFE are limited and hardware-costly~\cite{ozer:nature2024:bendableRiscV,Bleier:ISCA:2020:printedmicro}.
Additionally, folded architectures rely on conventional arithmetic units, such as multipliers--which are significantly larger than their bespoke counterparts~\cite{Mubarik:MICRO:2020:printedml}--thereby limiting the potential advantages of bespoke design.
Fully parallel architectures address these issues; however, their hardware overheads might scale with model size.

\subsubsection{Approximate Bespoke ML Classifiers}
\begin{figure}[t!]
    \centering
    \includegraphics[width=\linewidth]{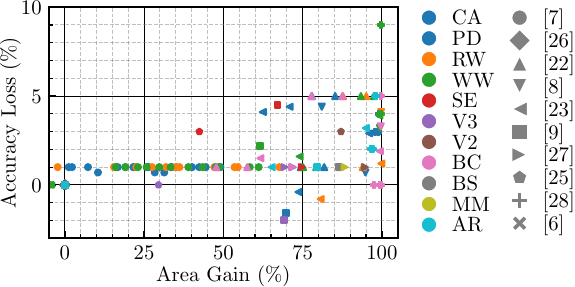}
    \caption{Accuracy-area trade-offs among existing approaches in PFE, across commonly-used datasets.}
    \label{fig:area_accuracy}
\end{figure}

While bespoke implementations offer high efficiency, additional optimization is essential for more complex classifiers, as hardware overheads remain prohibitively high for practical applications~\cite{Afentaki:ICCAD23:hollistic}.
Specifically targeting digital neural network classifiers, there has been a surge of research on designing approximate bespoke circuits, leveraging the perfect match formed between ML circuits and approximate computing.
An overview of the current state-of-the-art achievements in approximate printed neural networks' design is illustrated in Fig.~\ref{fig:area_accuracy}.

\begin{table}[ht]
  \centering
  \caption{Natively flexible digital bespoke ML classifiers}
  \label{tab:table_bespokeclf}
  \renewcommand{\arraystretch}{1.1}
  \setlength\tabcolsep{4pt}
  \begin{threeparttable}
    \begin{tabularx}{\columnwidth}{P{0.6cm}|c|P{1.2cm}|P{1.6cm}|P{1cm}|P{2cm}}
      \hline
      \textbf{Ref.} & \textbf{Tech.} & \textbf{ML Classifier} & \textbf{Approx.} & \textbf{Acc. Loss} & \textbf{Analog-Digital Interfacing}\\ \hline
      \cite{Ozer:Nature:2020} & FlexIC &  Custom GNB$^*$ & - & - & 5b ADCs \\\hline
      \cite{Ozer:DFLEPS2020} & FlexIC & BNN & - & - & 5b ADCs \\\hline
      \cite{velu24} & FlexIC & WNN & - & - & ADCs (est. $\geq$8b) \\\hline
      \cite{Duarte:ASPDAC2025} & FlexIC & MLP & In-train: pow2 weights \& Ax.ADC & $<$5\% & pruned 4b Binary ADCs \\\hline
      \cite{Mubarik:MICRO:2020:printedml} & EGFET & DT, SVM & - & - & Not Reported \\\hline
      \cite{Sertaridis:ISCAS2025} & EGFET & SVM & - & - & 4b ADCs \\\hline
      \cite{Armeniakos:DATE2022:axml} & EGFET & MLP, SVM & Post-train: Ax.Mult \& Gate-pruning & $<$2\% & 4b ADCs \\\hline
      \cite{Armeniakos:TCAD2023:cross} & EGFET & MLP, SVM & Post-train: Ax.Mult \& Gate-pruning \& VOS & $<$2\% & 4b ADCs \\\hline
      \cite{Armeniakos:TC2023:codesign} & EGFET & MLP & In-train: Ax.Mult Post-train: Ax.Add & $<$2\% & 4b ADCs \\\hline
      \cite{Balaskas:ISQED2022:axDT} & EGFET & DT & Post-train: Ax.Compare & $<$1\% & 4b ADCs \\\hline
      \cite{Kokkinis:TC2025} & EGFET & MLP & Neural Minimization & $<$2\% & 4b ADCs \\\hline
      \cite{Afentaki:ICCAD23:hollistic} & EGFET & MLP & In-train: pow2 weights Post-train: Ax.Add \& Ax.Relu & $<$5\% & 4b ADCs \\\hline
      \cite{Afentaki:DATE2024:gatrain} & EGFET & MLP & In-train: pow2 weights \& Ax.Add & $<$5\% & 4b ADCs \\\hline
      \cite{Mrazek:ICCAD2024:tnn} & EGFET & TNN & Post-train: Ax.Popcount & $<$1\% & 1b ADCs \\\hline
      \cite{Armeniakos:DATE2024:dt} & EGFET & DT & In-train: Ax.ADC & $<$1\% & pruned 4b Flash ADCs \\\hline
      \cite{Afentaki:ESL2024} & EGFET & MLP & In-train: pow2 weights \& Ax.ADC & $<$5\% & pruned 4b Flash ADCs \\\hline
    \end{tabularx}
    \begin{tablenotes}\footnotesize
      \item[$\star$] Gaussian Naïve Bayes.
    \end{tablenotes}
  \end{threeparttable}\vspace{-3ex}
\end{table}

The authors in~\cite{Armeniakos:DATE2022:axml} demonstrated that in bespoke neural networks, the classifier’s hardware overheads are closely related to the values of the model's coefficients.
To exploit this correlation,~\cite{Armeniakos:DATE2022:axml,Armeniakos:TCAD2023:cross} proposed a post-training approximation technique in which model weights are strategically replaced with hardware-friendly alternatives, enabling more area-efficient bespoke multipliers.
Specifically, replacing weights with nearby values (e.g., within $\pm4$) and balancing positive and negative replacements results in $28$\% area reduction for negligible accuracy loss~\cite{Armeniakos:DATE2022:axml}. 
To enable more aggressive hardware-friendly weight selection,~\cite{Armeniakos:TC2023:codesign} integrated this concept into the training process itself, incorporating the hardware cost of bespoke multipliers as a regularization term.
This co-design strategy optimizes both model accuracy and area efficiency.
Indicatively, for only up to $2$\% accuracy loss, $3.8$x lower area can be achieved compared to the exact bespoke design~\cite{Armeniakos:TC2023:codesign}.
\begin{figure}[h!]
    \centering
    \includegraphics[width=.75\linewidth]{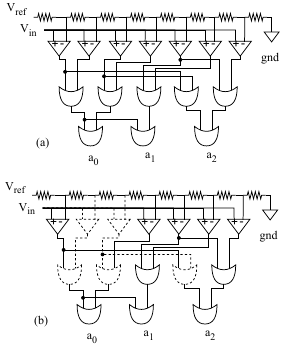}
    \caption{Schematics of (a) 3-bit Flash ADC~\cite{Afentaki:ESL2024}, and (b) its bespoke pruned counterpart~\cite{Duarte:ASPDAC2025}.}\vspace{-4ex}
    \label{fig:bespoke_adcs}
\end{figure}

\begin{figure*}[t]
\vspace{-1em}
\centerline{\includegraphics[width=1\linewidth]{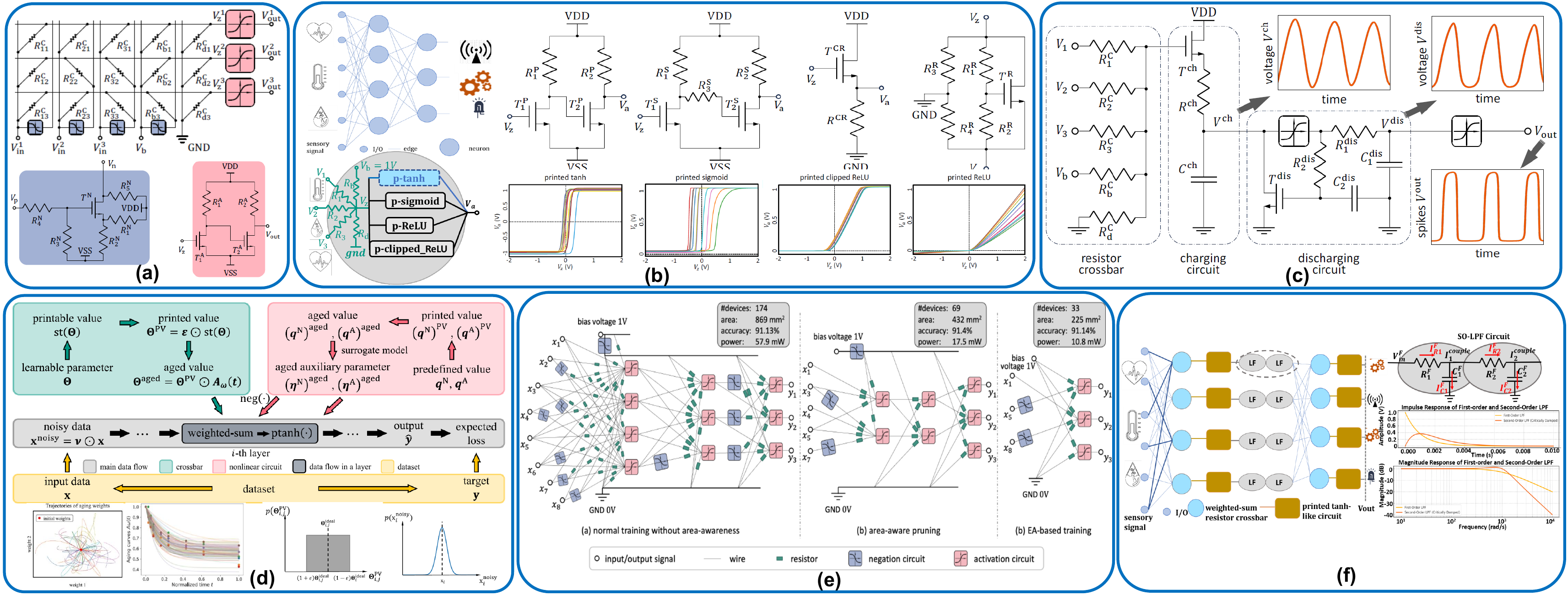}}
\caption{(a) Schematic of an exemplary printed neuron, negative weight circuit and tanh circuit. (b) Overview of neural architecture search using four different activation functions (tanh, sigmoid, clipped ReLU, ReLU). (c) Printed spiking neuromorphic circuit. (d) Dataflow in the dependability-aware training for pNCs. (e) Area-aware training and evolutionary-based training of a pNC. (f) Overview of an adaptive second-order low-pass filter based pNC for temporal data processing.} 
\label{fig:primitive}
\vspace{-1.5em}
\end{figure*}

Despite the significant efficiency gains in~\cite{Armeniakos:TC2023:codesign},~\cite{Afentaki:ICCAD23:hollistic} found that multipliers (even hardware-friendly) still require considerable area in many cases.
To address this,~\cite{Afentaki:ICCAD23:hollistic} leveraged the fact that in bespoke circuits, multiplication by a power of two can be implemented through simple rewiring.
By restricting model weights to power-of-two values,~\cite{Afentaki:ICCAD23:hollistic} completely eliminated the need for multipliers and further optimized the design by addressing the remaining bottleneck, i.e., additions, through adder-tree pruning.
Building on this idea,~\cite{Afentaki:DATE2024:gatrain} integrated all applied approximations directly into a genetic-based training process, in which a fine-grained bit-level pruning of input features is implemented.
Compared to the exact bespoke implementation, eliminating multipliers results in $2.5$x to $5$x area reduction for $1.25$\% average accuracy loss~\cite{Afentaki:ICCAD23:hollistic}, while the approximation-aware training achieves $181$x average area reduction with less than $5$\% accuracy loss~\cite{Afentaki:DATE2024:gatrain}.
Further maximizing hardware efficiency,~\cite{Mrazek:ICCAD2024:tnn} proposed a multiplier-less inference approach by utilizing only $1$-bit inputs and ternary weights while approximating the remaining additions through approximate popcount units generated via Cartesian genetic programming.
This approximation delivers a $2$x area reduction compared to~\cite{Afentaki:DATE2024:gatrain} while maintaining comparable accuracy.

Since most printed and flexible classification systems process sensor-based inputs, a significant portion of the system’s area may be occupied by the mandatory ADCs.
To address this, researchers extended the bespoke design paradigm to ADC design as well~\cite{Armeniakos:DATE2024:dt,Duarte:ASPDAC2025,Afentaki:ESL2024}.
Bespoke ADC design is implemented by keeping the bare minimum representations required for each sensor and eliminating redundant circuitry.
Examples of bespoke ADCs are shown in Fig.~\ref{fig:bespoke_adcs}.
At the cost of some accuracy degradation,~\cite{Armeniakos:DATE2024:dt, Duarte:ASPDAC2025,Afentaki:ESL2024} introduced in-training optimizations for fully-tailored ADCs, customized per input and model, resulting in substantial interfacing efficiency improvements. 
For example, for $1$\% accuracy loss, ADC costs in~\cite{Afentaki:ESL2024} are reduced by $8$x.

\subsection{Domain-specific Analog Processing Elements}\label{subsec:analog}

The development of domain-specific analog processing elements is a critical area within PFE, particularly for applications demanding low cost, flexibility, and biocompatibility that traditional silicon-based electronics often cannot meet. 
Flexible neuromorphic~\cite{Lebanov:TBCAS2024,Velazquez:Nature2024} and Printed neuromorphic circuits (pNCs) have emerged as a promising solution for such applications, drawing inspiration from the operational principles of artificial neural networks (ANNs) and spiking neural networks (SNNs) to perform computational tasks directly in the analog domain. 
This analog processing capability is particularly advantageous, as it eliminates the need for costly ADCs, thereby reducing device count and power consumption, which are critical constraints in many target applications.
These pNCs can be categorized into three key groups according to their specific functionalities and characteristics, as described below.

\subsubsection{Energy-Efficient pNCs}
\begin{figure}
    \centering
    \includegraphics[width=\linewidth]{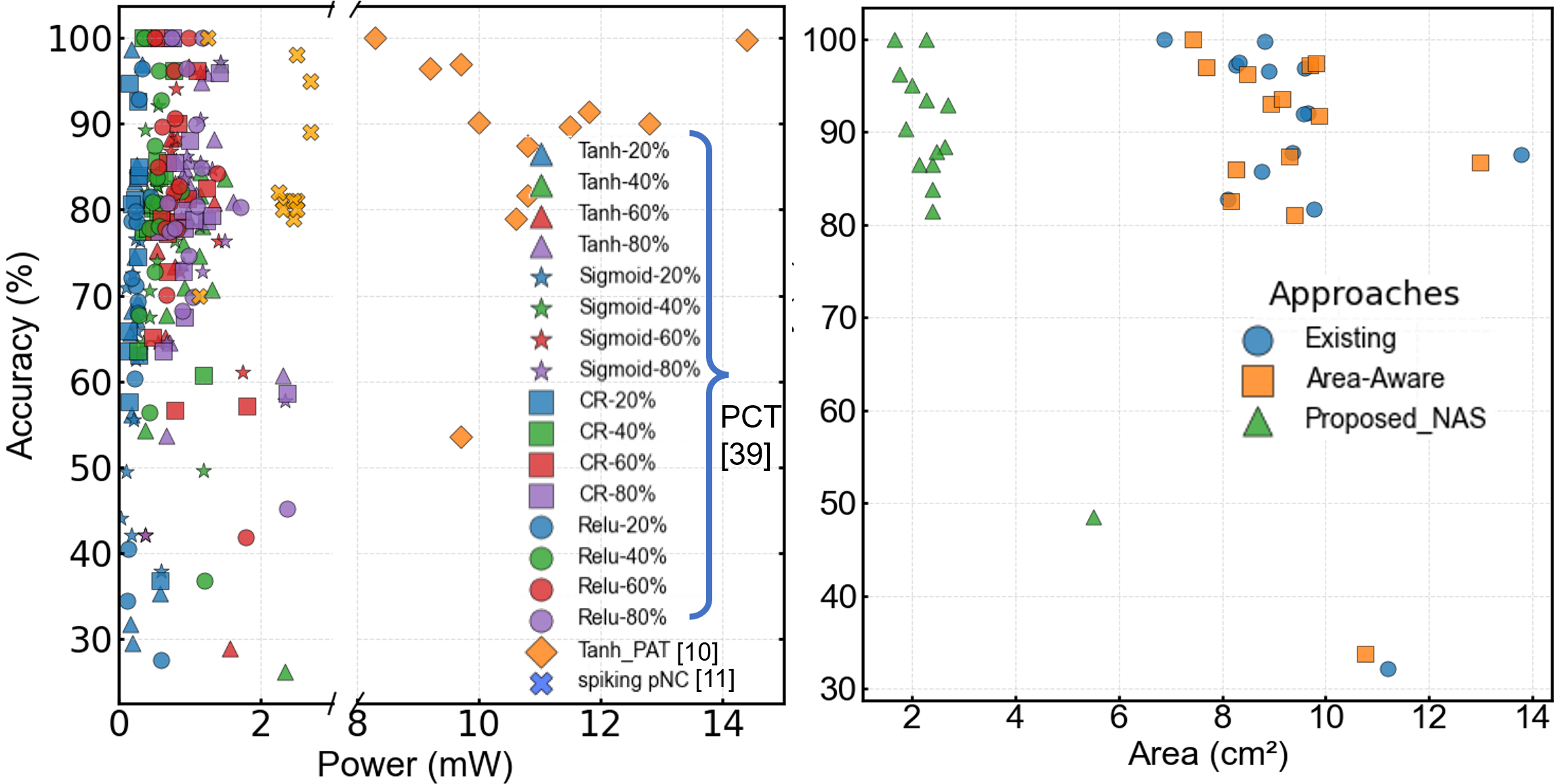}
    \caption{Accuracy-power-area trade-offs among existing analog pNCs for different datasets~\cite{dac_pnc,poweraware,spiking_pnc,nas_iccad,compact_pnc}.}
    \label{fig:energy_efficient_pNCs}\vspace{-5ex}
\end{figure}

These circuits are specifically designed for minimal power consumption, addressing the stringent power budgets typical of printed electronics applications. Resistor crossbars are employed to emulate weighted-sum operations, a core function in neural networks. As shown in Fig. \ref{fig:primitive} (a)-(b), tanh-like activation circuits and other analog nonlinear circuits implement essential nonlinear activation functions, allowing the processing of complex data patterns with minimal power consumption~\cite{poweraware}. Additionally, spiking neuromorphic circuits~\cite{spiking_pnc} (spiking pNCs, see Fig.~\ref{fig:primitive}(c)), which emulate biological neuron behavior by using discrete spikes to process information, significantly enhance energy efficiency.
Power-aware training methodologies are integrated into the design process to explicitly consider the energy consumption of individual circuit primitives, ensuring operation within strict power constraints\cite{dac_pnc} typically imposed by printed batteries or energy harvesting systems.
An overview of the current state-of-the-art achievements in area-power-accuracy tradeoff is illustrated in Fig.~\ref{fig:energy_efficient_pNCs}.

\subsubsection{Bespoke pNCs}
This category emphasizes highly customized designs, where the circuit architecture and component parameters are optimized specifically for particular datasets and application requirements. Such customization includes detailed parameterization of nonlinear components like activation functions and inverter-based circuits for negative weight operations\cite{lnc, nas_iccad} as shown in Fig.~\ref{fig:primitive}(b). Algorithms such as neural evolutionary architecture search~\cite{nas_iccad} and network pruning\cite{compact_pnc} (see Fig.~\ref{fig:primitive}(e)) are employed to achieve compact, area-efficient circuit designs, minimizing device counts and ensuring optimal performance tailored explicitly to the application's unique requirements.

\subsubsection{Temporal Processing pNCs}
\begin{figure}[t!]
    \centering
    \vspace{-1ex}
    \includegraphics[width=\linewidth]{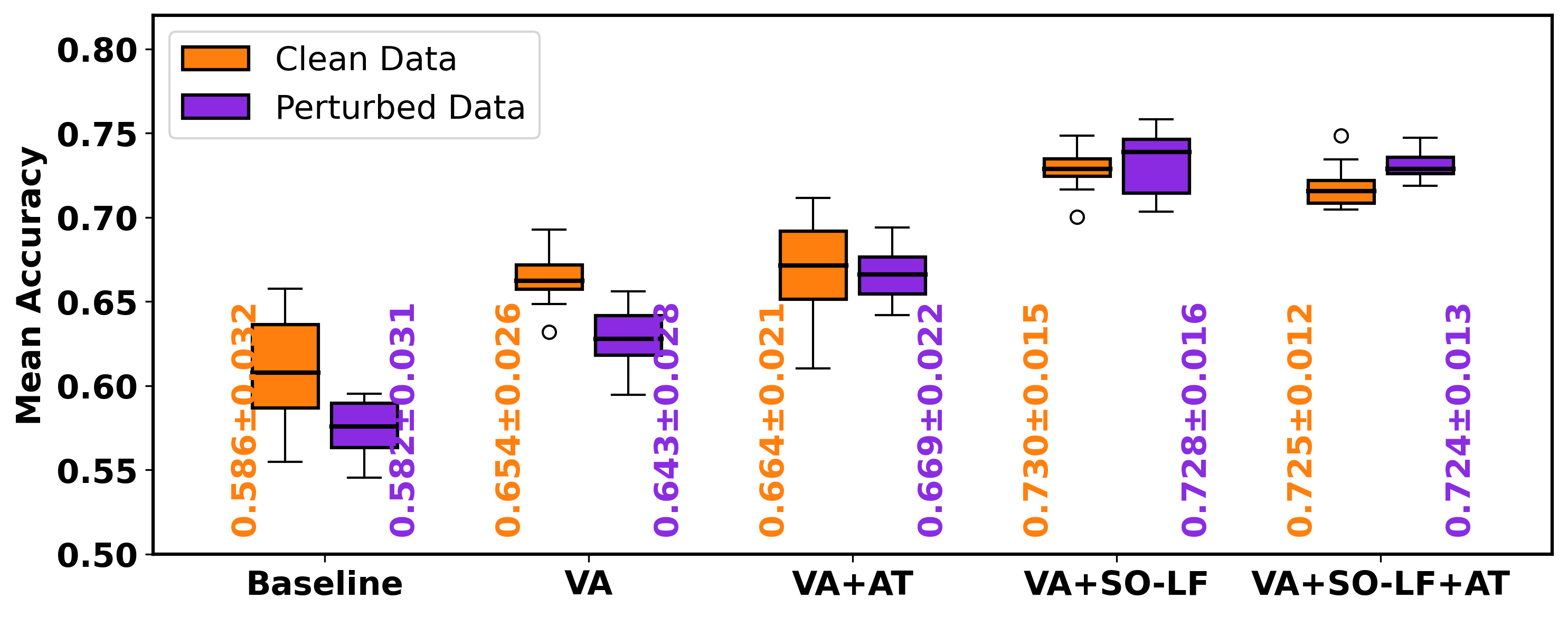}
    \caption{Average accuracy comparison between variation-aware (VA), augmented training (AT), second-order learnable filters (SO-LF), and VA+SO-LF+AT with baseline\cite{zhao2023towards} using an ablation study~\cite{adapt_pnc}.}
    \label{fig:adapt}\vspace{-2ex}
\end{figure}

These circuits as shown in Fig.~\ref{fig:primitive}(f), specialize in handling temporal sensory data by integrating learnable filters~\cite{zhao2023towards,adapt_pnc} into printed temporal processing blocks (pTPBs).
These blocks introduce time-dependent components, such as capacitors, to retain and process information from previous time steps. This temporal processing capability is crucial for applications like stress detection, where accurately capturing and analyzing temporal dynamics is essential. 
\cref{fig:adapt} presents an accuracy overview of temporal datasets~\cite{UCRArchive}, using an ablation analysis along with augmented training and second-order filters.  
Variation-aware training models process-level variations during circuit optimization, while ``clean'' and ``perturbed'' data refer to ideal and noise-augmented inputs used to evaluate robustness.

\section{Robustness and Reliability}
\label{sec:reliability}

To address the manufacturing variation and reliability challenges of pNCs,  various robustness-aware training methodologies have been developed that integrate the effects of manufacturing variations\cite{lnc,nas_iccad}, aging\cite{agingaware}, reliability\cite{nas_iccad,lnc} and sensing uncertainty\cite{Zhao:IDP:2023:variability_pe} into the design process.
As shown in Fig.~\ref{fig:adapt}, variation-aware training techniques aim to enhance the resilience of printed and flexible circuits by explicitly modeling and accounting for the statistical variations in component values during the training phase, allowing the network to learn parameters that are less sensitive to these variations. Similarly, aging-aware training methodologies focus on optimizing the initial parameters of pNCs by anticipating the degradation of their components over time, aiming to maintain acceptable levels of accuracy and performance throughout the expected lifetime of the device.

\begin{figure}[t!]
    \centering
    \includegraphics[width=\linewidth]{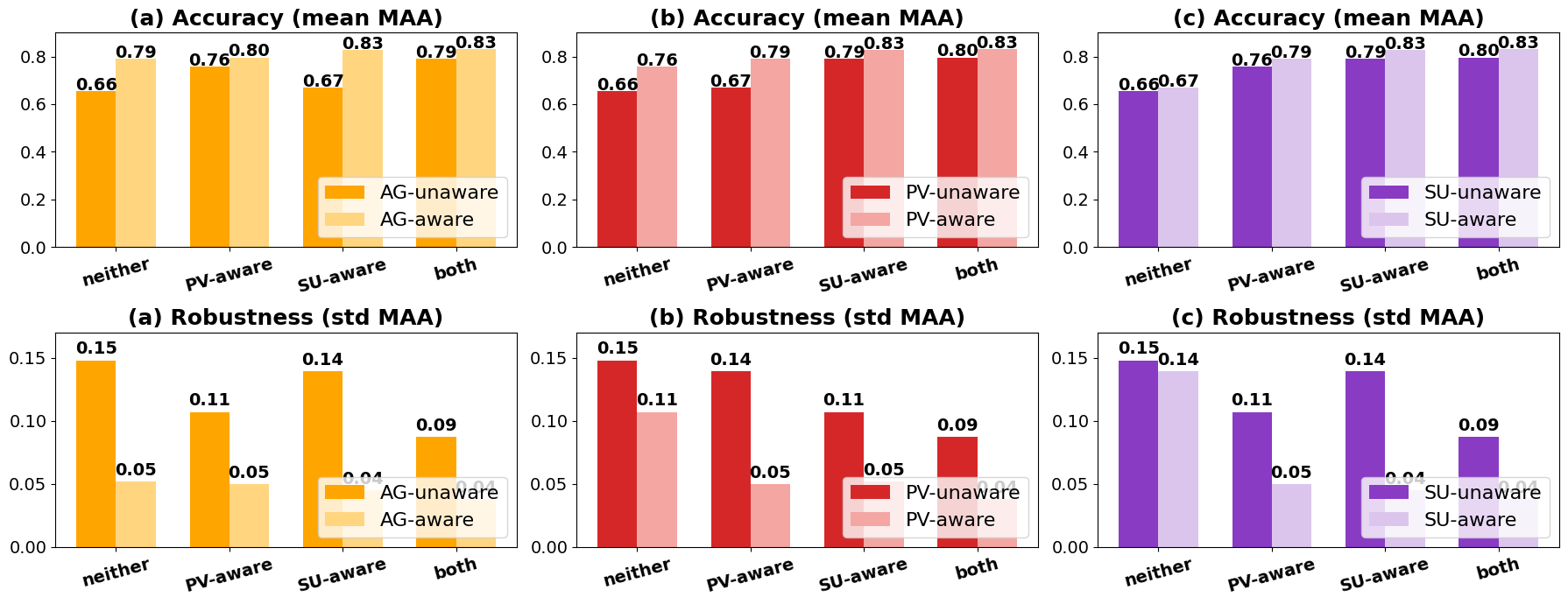}
    \caption{Combined effects of aging (AG), printing variation (PV), and sensing uncertainty (SU) on accuracy in the
dependability-aware training of pNCs. 
(MAA:measuring-aware accuracy)\cite{Zhao:IDP:2023:variability_pe}. }
    \label{fig:aging_pNCs}
\end{figure}
\begin{figure}[t!]
    \centering
    \includegraphics[width=\linewidth]{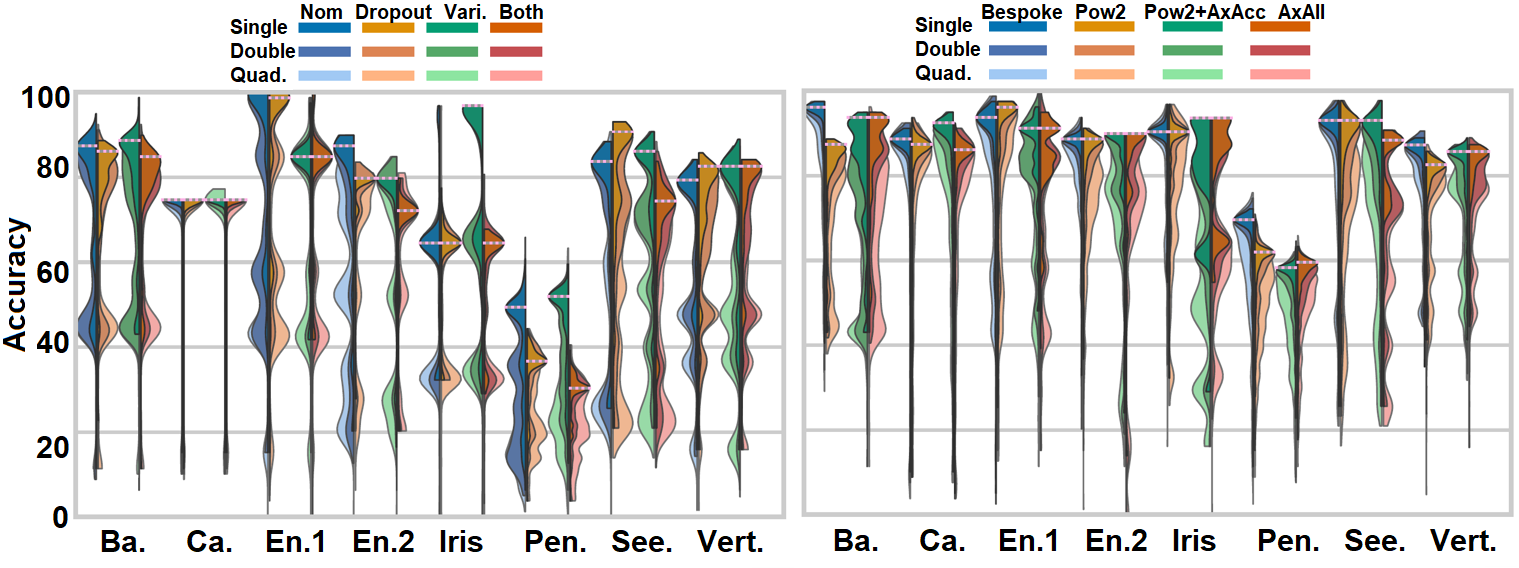}
    \caption{Evaluation of fault sensitivity w.r.t design approaches in analog pNCs and design architectures in digital pNCs\cite{fault_pnc}.}
    \label{fig:faulty_pNCs}
\end{figure}
\begin{table}[t!]
\setlength{\tabcolsep}{2pt}
\centering
\caption{Comparison of Gradient-based Algorithm and Random Search for automatic test input generation. 
}
\label{tab:results}
\begin{threeparttable}
\resizebox{\columnwidth}{!}{%
\begin{tabular}{|l|c c c|c c|c c|}
\hline
\multirow{3}{*}{\textbf{Dataset}}  & 
\multicolumn{3}{c|}{\textbf{Faults$^1$}} & 
\multicolumn{2}{c|}{
\begin{tabular}{@{}c@{}}\textbf{Gradient}\\\textbf{Based}\end{tabular}
} &
\multicolumn{2}{c|}{
\begin{tabular}{@{}c@{}}\textbf{Random}\\\textbf{Pattern}\end{tabular}
}
\\
\cline{2-8}

& \textbf{All} & 
\textbf{+Cluster} &
\begin{tabular}{@{}c@{}}\textbf{+Untest.}\\\textbf{Removal}\end{tabular} &
\begin{tabular}{@{}c@{}}\textbf{Fault}\\\textbf{Cov.$^2$}\end{tabular} &
\textbf{Vec$^3$} & 
\begin{tabular}{@{}c@{}}\textbf{Fault}\\\textbf{Cov.$^2$}\end{tabular} &
\textbf{Vec$^3$}  
 \\
\hline
\textbf{AcuteInfl} & 362 & 220 & 123 (2.94$\text{x}$) & 93.50\% & 115 & 47.96\% & 59 \\
\textbf{BreastCanc} & 434 & 265 & 128 (3.39$\text{x}$)  & 99.22\% & 127 & 49.22\% & 63 \\
\textbf{EnergyY2} & 432 & 267 & 214 (2.02$\text{x}$) & 97.66\% & 209 & 71.96\% & 154 \\
\textbf{Iris} & 336 & 207 & 154 (2.18$\text{x}$) & 97.40\% & 150 & 67.53\% & 104 \\
\textbf{Pendigits} & 778 & 506 & 448 (1.74$\text{x}$) & 98.21\% & 440 & 78.79\% & 32 \\
\textbf{Seeds} & 408 & 252 & 201 (2.03$\text{x}$) & 96.52\% & 194 & 72.64\% & 146 \\
\textbf{TicTacToe} & 434 & 265 & 153 (2.84$\text{x}$) & 86.27\% & 132 & 60.13\% & 92 \\
\textbf{VertCol2} & 362 & 220 & 180 (2.01$\text{x}$) & 69.44\% & 125 & 40.00\% & 72 \\
\textbf{VertCol3} & 384 & 237 & 124 (3.1$\text{x}$) & 99.19\% & 123 & 69.35\% & 86 \\
\hline
\textbf{Average} & \textbf{436.7} & \textbf{271.0} & \textbf{191.7 (2.3$\text{x}$)} & \textbf{93.05\%} & \textbf{179.4} & \textbf{61.95\%} & \textbf{89.83} \\
\hline
\end{tabular}%
}
\begin{tablenotes}\footnotesize
\item[] 
$^1$ Faults: Total faults before and after clustering, and after clustering + untestable removal,
$^2$ Fault Cov.: Fault Coverage (\%),
$^3$ Vec: Number of Test Vectors.
\end{tablenotes}
\end{threeparttable}
\vspace{-1.0em}
\label{atpg}
\end{table}

Dependability-aware training~\cite{Zhao:IDP:2023:variability_pe} represents a more comprehensive approach that simultaneously considers the impact of sensing uncertainty, printing variations, and aging, striving to design circuits that are robust against a combination of these factors, as shown in Fig.~\ref{fig:aging_pNCs}. Data augmentation techniques, including frequency domain augmentation, random cropping, jittering, time warping, and magnitude scaling, are also employed~\cite{adapt_pnc} (see Fig.~\ref{fig:adapt}) to improve the model's adaptability and robustness to noisy sensory inputs and physical variations encountered in real-world scenarios.
Moreover, understanding the impact of potential defects on the functionality of printed circuits is crucial for ensuring reliability.
Fault sensitivity analyses of both analog pNCs and digital pNCs are being performed at both circuit and algorithmic levels to evaluate how catastrophic faults in transistors and resistors (such as stuck-open and stuck-short faults) can affect their classification accuracy~\cite{fault_pnc}, as can be observed in Fig.~\ref{fig:faulty_pNCs}. 
To facilitate the detection of such faults and ensure reliable operation, an automatic test pattern generation (ATPG) framework has been developed, specifically tailored for pNCs~\cite{atpg_ets}. This framework aims to generate efficient test inputs that can maximize the output discrepancies between fault-free and faulty circuits, enabling comprehensive fault detection with a reduced number of test vectors. 
As shown in Table~\ref{atpg}, the gradient-based method significantly improves fault coverage and testing efficiency for analog pNCs, achieving over $90$\% fault coverage across datasets, outperforming random pattern testing. 
Additionally, we observe how fault abstraction significantly reduced the fault space (e.g., by 3.39× in \textit{BreastCanc}). However, datasets with complex fault behaviors, such as \textit{TicTacToe} and \textit{VertCol2}, remain challenging, often causing optimization to get stuck in local optima and thus require extra iterations for better coverage.
These collective efforts in robustness-aware design, fault analysis, and test generation are essential steps towards establishing the dependability of printed analog processing elements and enabling their widespread and reliable use in various emerging applications.

\begin{figure}[t!]
    \centering
    \includegraphics[width=\linewidth]{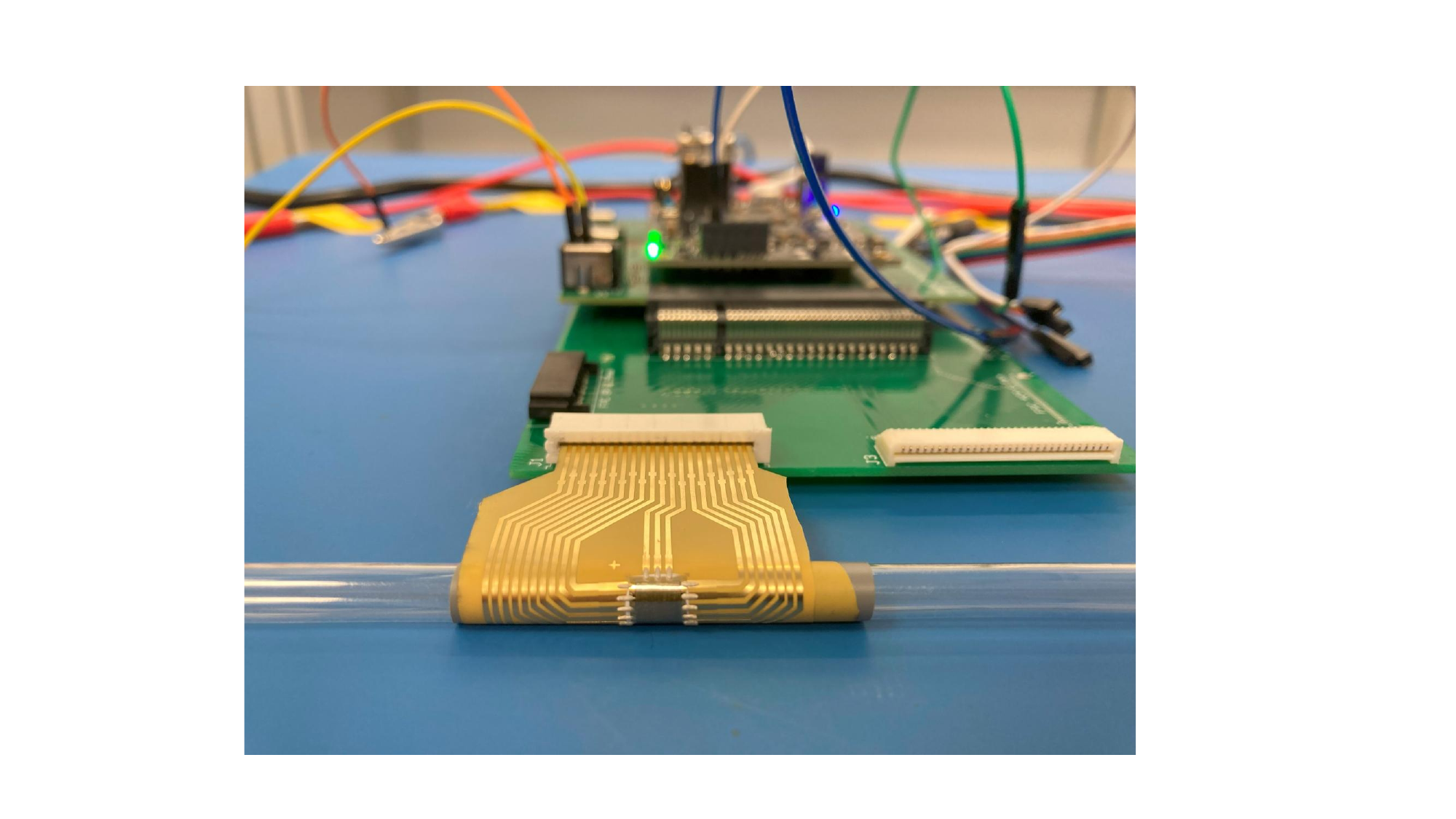}
    \caption{Flex-RV bent in a tensile mode while executing an application.}
    \label{fig:bent}\vspace{-2ex}
\end{figure}

Flexible electronic components endure significant mechanical stress when they are bent.
Thus, it is important to measure their mechanical strain in the form of bendability tests, in order to accurately assess their reliability and flexibility range in real-world deployment scenarios.
There have been many studies which perform parametric and functional tests at device level (TFT, electrodes etc.) under static and dynamic bending conditions~\cite{kim19, jang20, qazi24}.
The first study to demonstrate the bendability beyond device or a basic circuit level is Flex-RV~\cite{ozer:nature2024:bendableRiscV}, the 32-bit RISC-V processor fabricated as a FlexIC described in \cref{subsec:micpro}, which is assembled on a flexible PCB (FlexPCB). 
The bendability of Flex-RV under tensile and compressive strains has been validated through real application execution, as illustrated in Fig.~\ref{fig:bent}.
The results demonstrate that Flex-RV on a FlexPCV remains fully functional when bent to a 3 mm radius of curvature, exhibiting only minor performance variations depending on the applied bending strain.

\section{Conclusion}
Printed and flexible electronics offer a promising alternative to silicon-based technologies for low-cost, lightweight, and conformal applications, despite challenges in integration density, device variability, and circuit complexity.
Domain-specific circuits are enabled through extreme customizations--supported by the underlying technologies--while natively flexible general-purpose processors have been successfully fabricated and tested on flexible substrates and PCBs.
Reliability has been improved through variation- and aging-aware training, fault sensitivity analysis, and dedicated test generation.
Together, these developments show that printed and flexible systems can achieve competitive levels of hardware efficiency and robustness.
Continued algorithm-hardware co-design and System Technology Co-Optimization (STCO) for printed and flexible electronics remain essential to scale these technologies for practical, real-world deployment.

\section*{Acknowledgment}
This work is supported by the European Research Council (ERC) and co-funded by the H.F.R.I call “Basic Research Financing (Horizontal support of all Sciences)” under the National Recovery and Resilience Plan “Greece 2.0” (H.F.R.I. Project Number: 17048).


\end{document}